\documentclass[12pt,draftclsnofoot,onecolumn,journal,letterpaper]{IEEEtran}

\usepackage[final]{graphicx}
\usepackage[reqno]{amsmath}
\usepackage{amssymb}
\usepackage{url}
\usepackage{color}
\usepackage{times}
\usepackage{wrapfig}
\usepackage{subfig}
\usepackage{amsthm}
\usepackage{mathptmx}
\usepackage{amssymb}
\usepackage{amsfonts}
\usepackage{dsfont}
\usepackage{bm}
\usepackage{eucal}
\usepackage{mcite}
\usepackage{mathptmx}
\usepackage{multirow}
\usepackage{booktabs}
\usepackage{rotating}

\newtheorem{theorem}{Theorem}
\newtheorem{lemma}[theorem]{Lemma}
\newtheorem{remark}[theorem]{Remark}

\title{Haplotype Assembly: \\An Information Theoretic View}

\author{\IEEEauthorblockN{Hongbo~Si, Haris~Vikalo, and Sriram~Vishwanath\\}
\IEEEauthorblockA{
Department of Electrical and Computer Engineering\\
The University of Texas at Austin\\
Email: hongbosi@utexas.edu, \{hvikalo, sriram\}@ece.utexas.edu}}

\begin{document}

\maketitle


\begin{abstract}

This paper studies the haplotype assembly problem from an information theoretic perspective.
A haplotype is a sequence of nucleotide bases on a chromosome, often conveniently represented
by a binary string,
that differ from the bases in the corresponding positions on the other chromosome in a homologous
pair. Information about the order of bases in a genome is readily inferred using short reads provided
by high-throughput DNA sequencing technologies. In this paper, the recovery of the target pair of haplotype
sequences using short reads is rephrased as a joint source-channel coding problem. Two
messages, representing haplotypes and chromosome memberships of reads, are encoded and
transmitted over a channel with erasures and errors, where the channel model reflects salient features of high-throughput
sequencing. The focus of this paper is on the required number of reads for reliable haplotype reconstruction, and both the necessary and sufficient conditions are presented
with order-wise optimal bounds.

\end{abstract}


\section{Introduction}\label{sec:Introduction}

\begin{figure}[t!]
  \centering
  \includegraphics[width=1\columnwidth]{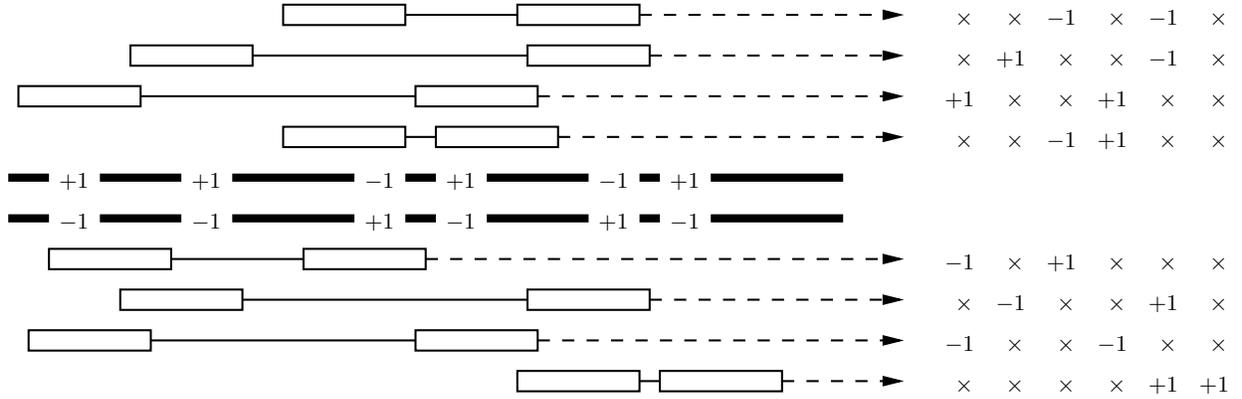}
  \caption{Paired-end reads generated from two chromosomes in a homologous pair. Rectangles linked
  by the lines above and below the target chromosome pair represent paired-end reads, and their relative
  positions indicate their location along the chromosomes.}\label{fig:Pairedend}
\end{figure}

Diploid organisms, including humans, have homologous pairs of chromosomes where one chromosome in
a pair is inherited from mother and the other from father. The two chromosomes in a pair are
similar and essentially carry the same type of information but are not identical. In particular, chromosomes
in a pair differ at a small fraction of positions (i.e., loci). Such variations are referred to as single nucleotide
polymorphisms (SNPs); in humans, frequency of SNPs is $1$ in $1000$. A haplotype is the string of SNPs on
a single chromosome in a homologous
pair. Haplotype information is essential for understanding genetic causes of various diseases and for
advancement of personalized medicine. However, direct measurement and identification of the entire
haplotype is generally challenging, costly, and time and labor intensive. Alternatively, single individual
haplotypes can be assembled from short reads provided by high-throughput sequencing systems.
These systems rely on so-called shotgun sequencing to oversample the genome and generate a
redundant library of short reads. The reads are mapped to a reference and the individual genome is
assembled following consensus of information provided by the reads.
The length of each read (i.e., DNA fragment) in state-of-the-art sequencing
systems is typically $100-1000$ base pairs \cite{schuster2007next}. Note that this length is comparable to
the average distance between SNPs on chromosomes. Therefore, a single read is unlikely
to cover more than one variant site which is needed for the haplotype assembly. Moreover, the origin of a
read (i.e., to which chromosome in a pair the read belongs) is unknown and needs to be inferred
\cite{schwartz2010assembly}. Paired-end sequencing \cite{campbell2008paired}, also known
as mate-paired sequencing \cite{edwards1990automated}, helps overcome these problems.
This process generates pairs
of short reads that are spaced along the target genome, where the spacing (so-called insert size) between
the two reads in a pair is known. The mate-pairs allow acquisition of the information about distant SNPs on
the same haplotype, and thus help assemble the haplotype. Fig.~\ref{fig:Pairedend} illustrates the procedure
of generating paired-end reads from a pair of chromosomes, where each read may cover two or more
variant sites. The goal of haplotype assembly is to identify the chromosome from which fragments are
sampled, and to reconstruct the haplotype sequences. When there are no sequencing errors, a fragment
conflict graph framework \cite{lancia2001conflict} converts the original problem into partitioning of the set
of reads into two subsets, each collecting the reads that belong to the same chromosome in a pair.
For erroneous data, it poses haplotyping as an optimization problem of minimizing the number of transformation steps needed to generate a bipartite graph \cite{lippert2002algorithmic}. This leads to
various formulations of the haplotype assembly problem including minimum fragment removal (MFR), minimum
SNP removal (MSR), and minimum error correction (MEC)\cite{lancia2001conflict}. The last one, MEC, has been the most
widely used criterion for haplotype assembly, due to its inherent relationship with independent error model.

In this paper, we analyze the haplotype assembly problem from information-theoretic perspective.
In particular, we determine necessary and sufficient conditions for haplotype assembly, both in the
absence of noise as well as for the case where data is erroneous. The paper is organized as follows.
Section II formalizes the haplotype assembly problem. In Section III, we present an information
theoretic view of haplotype assembly in the absence of sampling errors, and the erroneous case is
discussed in Section IV. Simulation results and analyses are shown in Section V. Finally, Section VI concludes the paper.


\section{Problem Formulation}\label{sec:Model}

A single nucleotide polymorphism (SNP) is the variation between two chromosomes where the
corresponding bases in a specific location on the chromosomes differ from each other. Typically,
diploid organisms have only two possible variants at one SNP site. For the sake of convenience,
we denote one of the two variants as $+1$ (i.e., the dominant one), while the other one we denote as $-1$ (i.e., the recessive one).
With this notation, a haplotype sequence $\bm{h}$ comprising information about all SNP sites on one of
the chromosomes in a pair can be represented by a string with elements in $\{+1,-1\}$, while the haplotype associated
with the other chromosome in the pair is its additive inverse $-\bm{h}$, where we denote
\begin{align}
&\bm{h}=(h_{1},h_{2},\ldots,h_{n}),\nonumber
\end{align}
and $n$ is the length of haplotypes (i.e., the number of SNPs).

Each paired-end read contains partial information about either of these two haplotypes.
Consider a set of discrete random variables $c_i$, where $i\in\{1,\ldots,m\}$ and $m$ is the number of reads.
Each $c_i$ corresponds to the chromosome membership for read $i$. More precisely, here,
\begin{align}
c_i=\left\{\begin{array}{ll}
+1,&\text{if read $i$ is sampled from $\bm{h}$,}\\
-1,&\text{if read $i$ is sampled from $-\bm{h}$.}
\end{array}
\right.\label{equ:c_i}
\end{align}
Due to the limitation of read lengths, only a
small fraction of entries are observed. In other words, a paired-end read $\bm{r}_i$ could be considered as a
sequence drawn from the alphabet $\{+1,-1,\times\}$, where ``$\times$'' refers to the lack of information
about that site. In the absence of sampling noise, every observed element $r_{ij}$ is obtained as the product
of the corresponding SNP and associated membership information \cite{puljiz2013message}.
Formally, this relationship is given by
\begin{equation}
r_{ij}=c_i\cdot h_{j}.\label{equ:Element_Relation}
\end{equation}

The collection of all reads forms a matrix $\bm{R}$, whose
rows correspond to $m$ paired-end reads, and whose columns correspond to $n$ SNP sites.
The $i$th row of $\bm{R}$ is denoted as $\bm{r}_i$ (i.e., read $i$), and the $j$th element of $\bm{r}_i$
is denoted as $r_{ij}$. Typically, only few entries in each row are numerical, if the effect of burst
variations is ignored.

By equation \eqref{equ:Element_Relation}, the observed matrix $\bm{R}$ could be interpreted as being obtained from a rank $1$ matrix
$\bm{S}$, whose row $\bm{s}_i$ is either $\bm{h}$ or $-\bm{h}$ based on the value of $c_i$, while most
of its entries are erased in the reading process. In particular, we have
\begin{equation}
\bm{R}=\mathcal{P}_{\bm{\Omega}}(\bm{S})\text{, where }\bm{S}=\bm{c}^T\cdot \bm{h},\label{equ:Rank_1}
\end{equation}
and $\bm{\Omega}$ is the collection of all observed locations, and the projection $\mathcal{P}$ is defined by
\begin{align}
\mathcal{P}_{\bm{\Omega}}(\bm{S})_{ij}=\left\{\begin{array}{ll}
s_{ij},&\text{if $(i,j)\in\bm{\Omega}$,}\\
\times,&\text{if $(i,j)\notin \bm{\Omega}$.}
\end{array}
\right.\label{equ:projection}
\end{align}
Hence, the task of haplotype assembly is to recover haplotype $\bm{h}$ and chromosome membership vector $\bm{c}$, or equivalently the matrix $\bm{S}$, from the observation matrix $\bm{R}$.

An example, illustrated by Fig.~\ref{fig:Pairedend}, corresponds to the scenario of $6$ SNP sites and $8$ paired-end reads.
Since only the first $4$ reads are (shotgun) sequenced from chromosome $1$, we obtain the chromosome membership vector $\bm{c}=(+1,+1,+1,+1,-1,-1,-1,-1)$.
If denoting the haplotype from chromosome $1$ as $\bm{h}=(+1,+1,-1,+1,-1,-1)$, then
the observed reads matrix, without the influence of error, is given by
\begin{align}
\bm{R}=\mathcal{P}_{\bm{\Omega}}(\bm{c}^T\cdot\bm{h})=\left[
\begin{array}{cccccc}
  \times & \times & -1 & \times & -1 & \times \\
  \times & +1 & \times & \times & -1 & \times \\
  +1 & \times & \times & +1 & \times & \times \\
  \times & \times & -1 & +1 & \times & \times \\
  -1 & \times & +1 & \times & \times & \times \\
  \times & -1 & \times & \times & +1 & \times \\
  -1 & \times & \times & -1 & \times & \times \\
  \times & \times & \times & \times & +1 & +1
\end{array}\right].\label{equ:Reads_Example}
\end{align}


\section{Error-free Case}

\begin{figure}[t!]
  \centering
  \includegraphics[width=1\columnwidth]{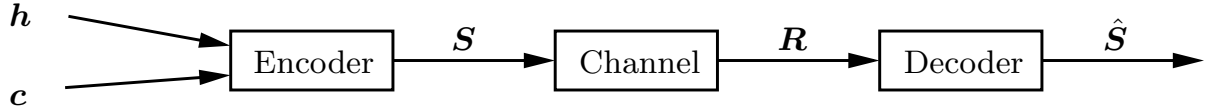}
  \caption{Information theoretic view of the haplotype assembly problem.}\label{fig:Model}
\end{figure}

From a joint source-channel coding perspective, haplotype assembly aims to recover two sources being
communicated through an erasure channel (see Fig.~\ref{fig:Model}). The first source is haplotype
information, $\bm{h}$, and the second source is the chromosome membership vector $\bm{c}$. Both of
these vectors are assumed to originate from a uniform distribution, i.e., their entries have $1/2$ probability to take values from $\{+1,-1\}$.
These two sources are encoded jointly using the function
$f:\{+1,-1\}^{n}\times\{+1,-1\}^m\to\{+1,-1\}^{m\times n}$, and the encoded codeword
$\bm{S}=f(\bm{h},\bm{c})$.  In particular, each entry in $\bm{S}$ is given by $s_{ij}=c_i\cdot h_{j}$,
which implies the encoder is a bijection.

After receiving the output from channel $\bm{R}$, the decoder uses the decoding function to map its channel observations into an estimate of the message.
Specifically, we consider the decoder (i.e., an algorithm for haplotype assembly) given by
$g:\{+1,-1,\times\}^{m\times n}\to \{+1,-1\}^{m\times n}$, such that $\hat{\bm{S}}=g(\bm{R})$, where $\hat{\bm{S}}$ represents the estimate.
Note that the encoding function is a bijection, decoding $\bm{S}$ is equivalent to decoding both $\bm{h}$ and $\bm{c}$.
We define the error probability of decoding as
\begin{equation}
P_{\text{e}}\triangleq \text{Pr}\{\hat{\bm{S}}\neq\bm{S}|\bm{R}\}.\label{equ:Error_Probability}
\end{equation}
As in the conventional information-theoretic analysis of a communication channel, we consider all
possible choices of matrix $\bm{S}$ (denote the resulting ensemble by $\mathcal{S}$), and let $m$ and $n$ be
large enough such that there exists at least one decoding function $g$ with small probability of error.

The channel model reflects particular reading technique. For the paired-end sequencing technique without sampling errors, let us consider the
channel $W:\{+1,-1\}^{m\times n}\to\{+1,-1,\times\}^{m\times n}$ described as follows:
\begin{enumerate}
\item Erasures happen independently across rows.
\item In each row, only $2$ entries remain and their positions are uniformly random.
\item Unerased entries are observed correctly.
\end{enumerate}
In other words, we observe precisely $2$ entries in each row of $\bm{S}$ for simpleness for the moment, and the observations are correct and
independent across different rows. Under these assumptions, the number of remaining entries in each column of $\bm{R}$
approximately obeys Poisson distribution. Moreover, the expected length of insert size between
$2$ sampled entries within a row is given by $(n-2)/3$. We should point out that the insert size in
practice is indeed limited and cannot be made arbitrarily large.

Based on this model for haplotype assembly without sequencing errors, we consider the necessary and sufficient conditions on the required number of reads for recovery.
\begin{theorem}\label{thm:ED}
Given $2$ arbitrary reliable observations in each row, the original haplotype matrix $\bm{S}$ could be reconstructed only if the number of reads satisfies
\begin{equation}
m=\Omega(n),\nonumber
\end{equation}
where $n$ is the length of target haplotype.
Moreover, if $m=\Theta(n\ln n)$, a reconstruction algorithm, erasure decoding, could determine $\bm{S}$ accurately with high probability. Specifically, given a target small constant $\epsilon>0$, there exists $n$ large enough such that by choosing $m=\Theta (n\ln n)$ the probability of error $P_{\text{e}}\leq \epsilon$.
\end{theorem}

We show the proof to necessary and sufficient conditions separately in the following two subsections.

\subsection{Necessary Condition for Recovery}

Using Fano's inequality \cite{Cover:IT1991}, we find that:
\begin{equation}
H(\bm{S}|\bm{R})\leq P_{\text{e}}\log|\mathcal{S}|\leq P_{\text{e}}(m+n),\label{equ:Fano_Inequality}
\end{equation}
where $\mathcal{S}$ is the assemble of all possible $\bm{S}$, and its size is upper bounded by $2^{m+n}$.

Recall $\bm{\Omega}$ comprises locations where $\bm{S}$ is observed, which is also random based on sampling locations.
Then, $\bm{\Omega}$ is independent of $\bm{S}$, and its rows are independent due to our channel assumption.
Therefore, we have
\begin{align}
H(\bm{S})   &\overset{(a)}{=} H(\bm{S}|\bm{\Omega})\nonumber\\
            &\overset{}{=}I(\bm{S};\bm{R}|\bm{\Omega})+H(\bm{S}|\bm{\Omega},\bm{R}) \nonumber\\
            &\overset{}{=}I(\bm{S};\bm{R}|\bm{\Omega})+H(\bm{S}|\bm{R}) \nonumber\\
            &\overset{(b)}{\leq}I(\bm{S};\bm{R}|\bm{\Omega})+P_{\text{e}}(m+n) \nonumber\\
            &\overset{}{=}H(\bm{R}|\bm{\Omega})-H(\bm{R}|\bm{S},\bm{\Omega})+P_{\text{e}}(m+n) \nonumber\\
            &\overset{(c)}{=}H(\bm{R}|\bm{\Omega})+P_{\text{e}}(m+n) \nonumber\\
            &\leq \sum_{i=1}^{m}H(\bm{r}_i|\bm{\omega}_i)+P_{\text{e}}(m+n) \nonumber\\
            &\overset{(d)}{=}2m+P_{\text{e}}(m+n) \nonumber
\end{align}
where $(a)$ follows from independence between $\bm{S}$ and $\bm{\Omega}$; $(b)$ from Fano's inequality, i.e., equation \eqref{equ:Fano_Inequality}; $(c)$ from the fact $\bm{R}$ is deterministic if $\bm{S}$ and $\bm{\Omega}$ are both known in the error-free case; and $(d)$ from the assumption that every row has exactly $2$ entries observed and noises are independent and symmetric.

Finally, by noting that $H(\bm{S})=m+n$, we need
\begin{align}
m\geq \frac{(1-P_{\text{e}})n}{1+P_{\text{e}}}.\label{equ:Lower_Bound}
\end{align}
for accurate recovery. More precisely, roughly we need $m=\Omega(n)$ for recovery with arbitrary small probability of decoding error.

\begin{remark}
Note that in this proof, channel model is only utilized when bounding $H(\bm{R}|\bm{\Omega})$. To this end, the necessary result is extendable to more channel models (i.e., reading techniques). In particular, the lower bound $m=\Omega(n)$ fits for deterministic choice of reading sites, paired-end reading with fixed insert size, and more importantly, reading techniques with more than $2$ observations in each read. The essential condition for the establishment of necessary condition is to ensure the matrix is sparse. More precisely, the number of all observed entries in the matrix is in proportion to $n$.
\end{remark}

\subsection{Sufficient Condition for Recovery}

The goal of a decoding algorithm is to recover $\bm{S}$ (or equivalently $\bm{h}$ and $\bm{c}$) from $\bm{R}$ with high confidence.
Here, we show a simple and effective algorithm, called ``erasure decoding'', which requires only $\Theta(n\ln n)$ number of reads for reliable haplotype recovery.
Detailed steps of this algorithm are described as follows:

\begin{enumerate}
\item Choose the ``seed'' $s$ as an arbitrary non-erased entry in the first row, i.e. $s=r_{1j}$, where $j$ is randomly chosen such that $r_{1j}\neq \times$.
Evaluate the membership of first row as $c_1=+1$.
\item Find all other rows with position $j$ not erased, i.e.,
\begin{equation}
\mathcal{A}=\{k|r_{kj}\neq \times,\; k\neq 1\}.
\end{equation}
\item Evaluate the membership of all rows with indices in $\mathcal{A}$ as
\begin{equation}
c_k=\left\{\begin{array}{ll}
+1, & \text{if }r_{kj}=r_{1j},\\
-1, & \text{otherwise},
\end{array}\right.
\end{equation}
for every $k\in\mathcal{A}$.
\item Decode SNPs in the first row by
\begin{equation}
r_{1l}=c_k\cdot r_{kl},
\end{equation}
for every $k\in\mathcal{A}$ and $r_{kl}\neq \times$.
\item Delete all rows with indices in $\mathcal{A}$.
\item Arbitrarily choose another non-erased entry in the first row as the new seed $s=r_{1j}$, which has not been chosen as seed in any of the former steps. Repeat Step $2)$ to $6)$ until no row could be further erased.
\item If the first row is the only remaining one and its entries are all decoded, claim $\bm{h}=\bm{r}_1$; otherwise claim a failure.
\end{enumerate}

\begin{remark}
In this algorithm, we arbitrarily evaluate a chromosome membership for the first row, but it may not be the correct one. In fact, if the algorithm successfully decodes both $\bm{h}$ and $\bm{c}$, then all elements could be flipped due to an incorrect choice of initial membership. However, the recovered matrix $\bm{S}$ remains the same, due to the particular product operation to generate $\bm{S}$. At this point, the choice of initial membership does not influence the decoding performance.
\end{remark}

\begin{remark}
Erasure decoding is closely connected to the bipartite partition interpretation \cite{schwartz2010assembly}. Note that if our algorithm successfully recovers the message matrix $\bm{S}$, we can realign its rows such that the matrix could be partitioned into two sub-matrices with different chromosome memberships. To this end, the erasure decoding provides a practical algorithm to fulfill partition a bipartite for haplotype assembly, in the error-free case.
\end{remark}

\begin{figure}[t!]
  \centering
  \includegraphics[width=1\columnwidth]{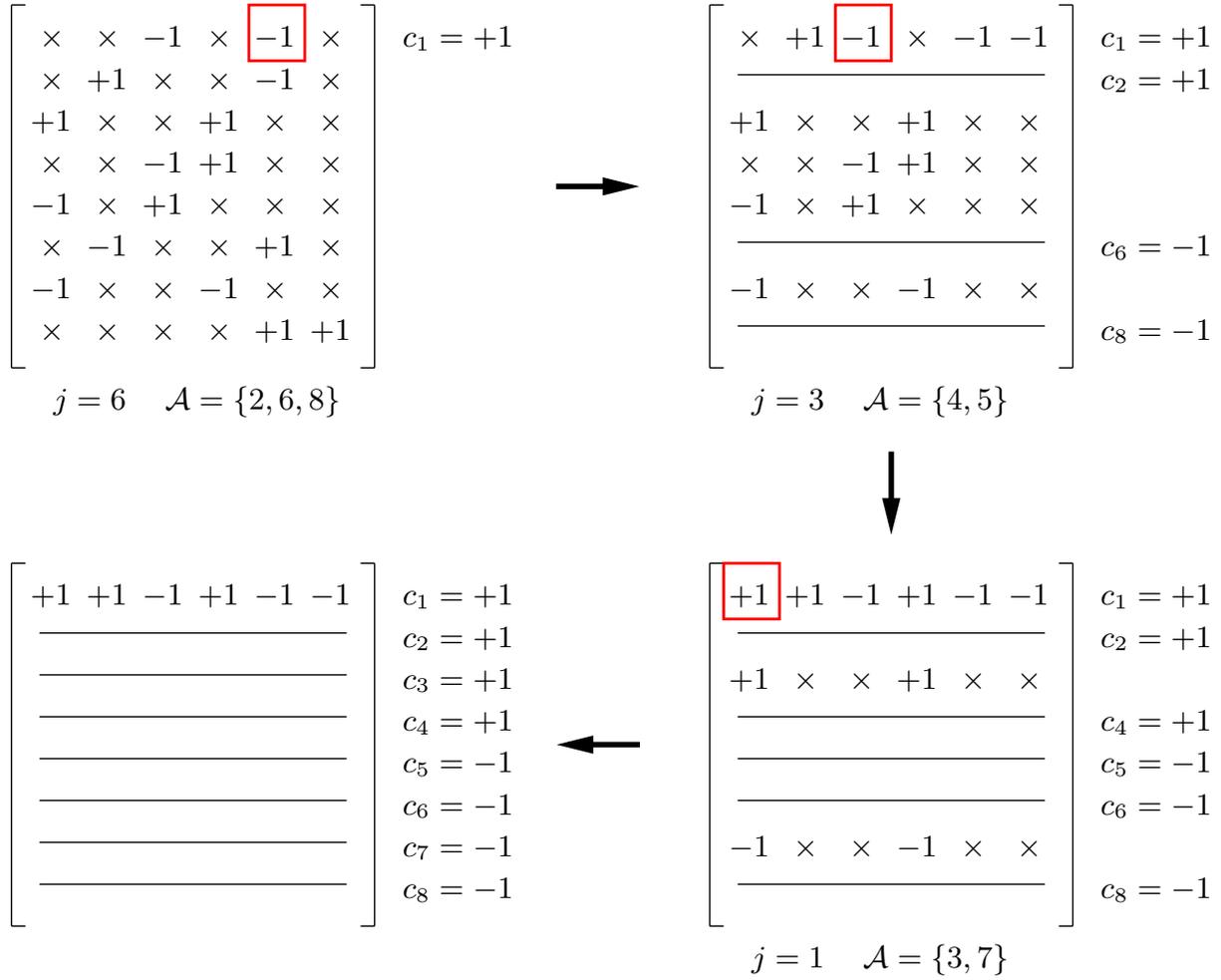}
  \caption{Erasure decoding of the example illustrated in Fig.~\ref{fig:Pairedend}. In every round (Step $2)$ to $6)$), the seed is marked in a rectangle, with its column index given by $j$. Rows that share the same positions observed as the seed are collected in assemble $\mathcal{A}$. A straight line crossing a whole row of the matrix represents a deletion. }\label{fig:Decoding}
\end{figure}

Fig.~\ref{fig:Decoding} shows the details of decoding procedures for the example illustrated in Fig.~\ref{fig:Pairedend}, where the read matrix is given by \eqref{equ:Reads_Example}.

Here, we analyze the performance of this proposed algorithm. More precisely, we show that if the number of reads sample is large enough, i.e. $m=\Theta(n\ln n)$, the source matrix $\bm{S}$ could be recovered correctly with high probability. Observe that in the absence of sampling errors, the erasure decoding algorithm ensures the output to be the correct haplotype if both of the following conditions are satisfied.
 \begin{enumerate}
 \item All rows except for the first one are deleted.
 \item All entries in the first row are decoded
 \end{enumerate}
At this point, decoding error occurs if at least one of the following events happen.
\begin{enumerate}
\item The event $E_1$: at least one of the columns in $\bm{R}$ is erased such that the corresponding
SNP could not be decoded;
\item The event $E_2$: there exist a partition of row indices $\{1,\ldots,m\}=\mathcal{U}_1 \cup \mathcal{U}_2$, and a partition of column indices $\{1,\ldots,n\}=\mathcal{V}_1 \cup \mathcal{V}_2$, such that $|\mathcal{V}_1|\geq 2$ and $|\mathcal{V}_2|\geq 2$ (to make sure $2$ entries could be sampled from each row), and $r_{ij}=\times$ for any $(i,j)\in\left(\mathcal{U}_1\times\mathcal{V}_2\right)\cup \left(\mathcal{U}_2\times\mathcal{V}_1\right)$. In other words, the sampled entries could be considered as originated from two disjoint subsets of target haplotypes and thus there is no hope to recover it due to the lack of information bridging these subsets.
\end{enumerate}

We outline how to bound the probability of each of the two error events. First, note that by coupon
collector effect, if $m=\Theta(n\ln n)$ then every column is covered by at least one read with high
probability. More precisely, by taking $m=n\ln n$, the error event (or equivalently the tail distribution for
coupon collector problem) is given by
\begin{align}
\text{Pr}\{E_1\}    &=\frac{\sum\limits_{i=1}^{n-2}{n\choose i}{{n-i}\choose 2}^m}{{n\choose 2}^m}\nonumber\\
                    &=\sum\limits_{i=1}^{n-2}{n\choose i}\left[\frac{(n-i)(n-i-1)}{n(n-1)}\right]^m\nonumber\\
                    &\leq \sum\limits_{i=1}^{n-2}n^i e^{-m\frac{2i n-i(i+1)}{n(n-1)}}\nonumber\\
                    &=\sum\limits_{i=1}^{n-2}O(n^{-i})\nonumber\\
                    &=O(n^{-1}). \label{equ:Error_1}
\end{align}

On the other hand, the second error event $E_2$ could be further decomposed into sub-events $E_2^{u,v}$ which represent the type $2$ error event with particular $u=|\mathcal{U}_1|$ and $v=|\mathcal{V}_1|$. Then, we have
\begin{align}
\text{Pr}\{E_2^{u,v}\}  &=\frac{{n\choose v}{m\choose u}{v\choose 2}^u{n-v\choose 2}^{m-u}}{{n\choose 2}^m}.\label{equ:Error_2}
\end{align}
Observe that by symmetry and monotonicity, the right hand side in \eqref{equ:Error_2} is maximized by two extreme points on the feasible $(u,v)-$region, i.e., for any $u$ and $v$, $\text{Pr}\{E_2^{u,v}\}\leq \text{Pr}\{E_2^{1,2}\}=\text{Pr}\{E_2^{m-1,n-2}\}$. In particular, we have
\begin{align}
\text{Pr}\{E_2^{1,2}\}  &=\frac{{n\choose 2}{m\choose 1}{2\choose 2}^1{n-2\choose 2}^{m-1}}{{n\choose 2}^m}\nonumber\\
                        &=\frac{m[(n-2)(n-3)]^{m-1}}{[n(n-1)]^{m-1}}\nonumber\\
                        &\leq n\ln n \left(1-\frac{4n-6}{n(n-1)}\right)^{n\ln n-1}\nonumber\\
                        &\leq n\ln n e^{-\frac{4n-6}{n(n-1)}(n\ln n-1)}\nonumber\\
                        &=O(n^{-3}\ln n).\nonumber
\end{align}
Hence, the probability of the second error event is upper bounded by
\begin{align}
\text{Pr}\{E_2\}&=\sum_{u=1}^{m-1}\sum_{v=2}^{n-2}\text{Pr}\{E_2^{u,v}\}\nonumber\\
                &\leq (m-2)(n-4)\text{Pr}\{E_2^{1,2}\}\nonumber\\
                &\leq n^2\ln n O(n^{-3}\ln n)\nonumber\\
                &=O(n^{-1}(\ln n)^2).\label{equ:Error_2_2}
\end{align}
Combining these two bounds together, we obtain
\begin{align}
P_{\text{e}}\leq \text{Pr}\{E_1\}+\text{Pr}\{E_2\}=O(n^{-1})+O(n^{-1}(\ln n)^2)< \epsilon, \nonumber
\end{align}
for arbitrary $\epsilon>0$ with sufficiently large $n$.

\begin{remark}
Note that there is a log-factor gap between the lower and upper bounds. As analyzed in \cite{vishwanath2010information}, this log-factor generally exists and ensures enough entries sampled from each column for accurate recovery. If a more systematic reading method could be adopted to generate the observation matrix, the log-factor may not be essential for reconstruction. We will see in the next section, for the erroneous case, this log-factor gap between two bounds also exists.
\end{remark}


\section{Erroneous Case}

When determining the SNP on a particular location, we basically perform a hypothesis testing between dominant and recessive.
To this end, when sequencing errors are present, some of the entries observed in $\bm{R}$ are flipped. Here, we
assume errors are independent and identically distributed.
More precisely, generation of errors could be modeled as messages passing through a set of
independent binary symmetric channels with parameter $p$, where $p$ is the probability of flipping the sign.
Denoting the noise as matrix $\bm{N}$, where $n_{ij}$ are i.i.d. distributed, we have
\begin{align}
\bm{R}=\mathcal{P}_{\bm{\Omega}}(\bm{S}+\bm{N}).\label{equ:Noise_Model}
\end{align}
Hence, the system model for the erroneous case could be considered as the one for error-free case concatenated with a channel representing the generation of noises. More precisely, the equivalent channel model $W:\{+1,-1\}^{m\times n}\to\{+1,-1,\times\}^{m\times n}$ considered in this section is described as follows:
\begin{enumerate}
\item Erasures happen independently across rows.
\item In each row, only $2$ entries remain and their positions are uniformly random.
\item Unerased entries have probability $p$ to be read incorrectly, and the errors happen independently.
\end{enumerate}

Then, for perfect recovery, we want to reconstruct $\bm{S}$ from $\bm{R}$ with high probability. However, if no more than two entries could be observed in a row, the identification of origin is not always feasible. For instance, if we observe $(+1,+1)$ in a particular read, and know only one error happens when sequencing. Then, there is no hope to discover whether the potential true sequence should be $(-1,+1)$ or $(+1,-1)$, although either way does not influence the decoding of haplotypes. To this end, in the erroneous case, we aim to recover the row space only, i.e., the haplotype $\bm{h}$, from matrix $\bm{R}$ with high probability. More precisely, if denoting the estimate from a recovery algorithm as $\hat{\bm{h}}$, we define the probability of error in this case as
\begin{equation}
P_{\text{e}}=\text{Pr}\{\hat{\bm{h}}\neq \bm{h}|\bm{R}\},\nonumber
\end{equation}
to evaluate the recovery accuracy. We desire this probability to be arbitrarily small, on an average across all possible implementation of $\bm{h}$.

Based on this model for haplotype assembly, we consider the necessary and sufficient conditions on the number of required reads for recovery.
\begin{theorem}\label{thm:SP}
Given $2$ arbitrary unreliable observations in each row, the original haplotype vector $\bm{h}$ could be reconstructed only if the number of reads satisfies
\begin{equation}
m=\Omega(n),\nonumber
\end{equation}
where $n$ is the length of target haplotype.
Moreover, if $m=\Theta(n\ln n)$, a reconstruction algorithm, spectral partitioning, could determine $\bm{h}$ accurately with high probability. Specifically, given a target small constant $\epsilon>0$, there exists $n$ large enough such that by choosing $m=\Theta (n\ln n)$ the probability of error $P_{\text{e}}\leq \epsilon$.
\end{theorem}

The theorem shows that although observations are not reliable due to sampling noises, the number of reads needed remains the same scale of $n$. Here, we provide details to show the proofs to both conditions separately in the following subsections.

\subsection{Necessary Condition for Recovery}

To study the necessary condition, one may still rely on the Fano's equality, i.e.,
\begin{align}
H(\bm{h}|\bm{R})\leq P_{\text{e}}\cdot n.\nonumber
\end{align}
Using this, we obtain
\begin{align}
H(\bm{h})\overset{}{\leq}H(\bm{R}|\bm{\Omega})-H(\bm{R}|\bm{h},\bm{\Omega})+P_{\text{e}}\cdot n.\nonumber
\end{align}
In this case, $H(\bm{R}|\bm{h},\bm{\Omega})$ does not vanish due to the influence of noise. In particular, by noting noises are assumed to be i.i.d., we have
\begin{align}
H(\bm{R}|\bm{h},\bm{\Omega})\geq H(\bm{R}|\bm{S},\bm{\Omega}) =\sum_{i=1}^mH(\bm{r}_i|\bm{s}_i,\bm{t}_i)=2mH(p).\nonumber
\end{align}
Combining with the observations that $H(\bm{R}|\bm{\Omega})\leq2m$ and $H(\bm{h})=n$, we have
\begin{equation}
m\geq \frac{(1-P_{\text{e}})n}{2[1-H(p)]},
\end{equation}
which is still an $m=\Omega(n)$ scale lower bound.

\subsection{Sufficient Condition for Recovery}

On the other hand, from the perspective of sufficient condition, if errors happen, erasure decoding
algorithm may not apply. With any luck, the algorithm sometimes recovers haplotypes correctly, but more commonly, erroneous items in the matrix may cause a failure.
In fact, an effective algorithm for haplotype assembly from a small number
of reads remains open. Most state-of-the-art algorithms are still based on graphical interpretation with an optimization formulation,
adopting different objective criteria \cite{lancia2001conflict}.

In particular, among these criteria, widely adopted ones include minimum fragment removal (MFR), minimum
SNP removal (MSR), and minimum error correction (MEC).
MFR \cite{lancia2001conflict} criterion aims to remove the minimum number of fragments (i.e., reads) so as to leave a bipartite.
The remaining graph is conflict-free and algorithms for error-free case could therefore be performed on it to recover the haplotypes.
However, to solve this optimization problem itself is not immediate, since it is non-convex in general.
MSR \cite{lancia2001conflict} criterion is an alternative formulation for the problem. Precisely, it removes the least number of SNP sites such that
the remaining graph could be partitioned into two haplotypes. From the perspective of graphical interpretation,
MSR aims to find the maximum independent set of the original graph.
MEC \cite{lippert2002algorithmic} criterion requires to flip the minimum number of entries in observed matrix to allow the assembly of haplotypes.
It corresponds to a straight-forward error-correction insight with i.i.d. noise generation model. To this end, MEC is the most widely preferred criterion
in studies for the moment. HapCUT \cite{bansal2008hapcut} is a typical algorithm to solve the MEC optimization problem.

However, in general, current algorithms basically consider the number of reads as a known parameter
for complexity analysis, rather than regarding it as the essential measurement to argue
sufficient condition for haplotyping. As a contrast, in this paper, we focus on the information theoretical view: the condition for perfect recovery. In particular, we propose a low-rank matrix interpretation for haplotype assembly. Intuitively, for haplotype assemble, we aim to partition all SNP sites into two sets, corresponding to dominant and recessive correspondingly. By regarding the adjacent matrix of the original graph as a perturbation of a planted model, which is a low rank matrix in nature, we claim that the partition is perfect as long as the parameters are chosen properly. In detail, first, we describe the ``spectral partitioning'' algorithm using SVD technique to obtain a weaker conclusion that the fraction of partition errors vanishes as $n$ increases, and then, we use a remark to discuss a modified algorithm for perfect recovery at the end of this section. The steps for spectral partitioning are described as follows:

\begin{enumerate}
\item Construct an adjacent matrix $\bm{A}\in\{0,1\}^{n\times n}$ based on the observation matrix $\bm{R}$, such that for every $(u,v)\in\{1,\ldots,n\}\times\{1,\ldots,n\}$ with $u>v$,
\begin{align}
a_{uv}=\left\{\begin{array}{ll}
1,&\quad \text{if }\sum\limits_{i=1}^{m}\bold{1}_{\{r_{iu}\neq\times,r_{iv}\neq\times,r_{iu}=r_{iv}\}}>\sum\limits_{i=1}^{m}\bold{1}_{\{r_{iu}\neq\times,r_{iv}\neq\times,r_{iu}\neq r_{iv}\}},\\
0,&\quad\text{otherwise.}
\end{array}
\right.\label{equ:Majority_Voting}
\end{align}
Then, let $a_{uv}=a_{vu}$ for any $u>v$ to guarantee symmetry, and let $a_{uu}=0$ for diagonal entries.
\item Perform singular value decomposition (SVD) to matrix $\bm{A}$, i.e. $\bm{A}=\bm{U}\bm{\Lambda}\bm{V}$ such that $\bm{U}, \bm{V}\in\mathbb{R}^{n\times n}$ are unitary matrices, and $\bm{\Lambda}\in\mathbb{R}^{n\times n}$ is diagonal.
\item Take the eigenvector $\bm{v}_2(\bm{A})$ corresponding to the second largest eigenvalue of $\bm{A}$, then construct sets
\begin{align}
\mathcal{C}_1=\{j:v_{2j}<0\}, \quad \mathcal{C}_2=\{j:v_{2j}\geq0\}.\nonumber
\end{align}
Based on this, the haplotype is recovered by
\begin{equation}
h_j=\left\{\begin{array}{ll}
+1, &\quad\text{if }j\in\mathcal{C}_1,\nonumber\\
-1, &\quad\text{if }j\in\mathcal{C}_2.\nonumber
\end{array}
\right.\nonumber
\end{equation}
\end{enumerate}

\begin{remark}
In words, for every entry in $\bm{A}$, \eqref{equ:Majority_Voting} performs a majority voting among all reads covering the corresponding SNP sites. Note that this is equivalent to MAP hypothesis testing with uniform prior distribution. Hence, if the distribution of haplotype is not assumed to be uniform, or error distributions are not identical across SNP sites, weighted majority voting should be utilized for a general case.
\end{remark}

Here, we analyze the performance of spectral partitioning by showing its relationship to the classical partitioning problem on a planted model. The intuition originates from the perturbation theory for eigenvectors, with respect to a low rank matrix. The analysis follows steps similar to \cite{mcsherry2001spectral}, but focusing on the particular haplotype assembly background such that bounds in this paper are much tighter.

\subsubsection{Planted Model}
First of all, we consider the planted model, i.e., a matrix $\bm{B}\in\mathbb{R}^{n\times n}$ defined as
\begin{align}
\bm{B}=\left[\begin{array}{cc}
\left[\alpha\right]_{n_1\times n_1} & \left[\beta\right]_{n_1\times n_2}\nonumber\\
\left[\beta\right]_{n_2\times n_1} & \left[\alpha\right]_{n_2\times n_2} \nonumber
\end{array}
\right],\nonumber
\end{align}
where $\alpha>\beta>0$, $n_1+n_2=n$, and $[\alpha]_{n_1\times n_1}$ represents for an $n_1\times n_1$ sub-matrix with all entries as $\alpha$.

By this particular construction, matrix $\bm{B}$ is believed to be low-rank. In fact, if we perform SVD on $\bm{B}$, it is evident to see the rank of $\bm{B}$ is $2$, and its first two singular values and corresponding singular vectors are given by
\begin{align}
\lambda_1(\bm{B})&=n_1\beta\mu_1+n_2\alpha\;,\label{equ:bar_lambda_1}\\
\lambda_2(\bm{B})&=n_1\beta\mu_2+n_2\alpha\;,\label{equ:bar_lambda_2}\\
\bm{v}_1(\bm{B})&=\left(\left[\frac{\mu_1}{\sqrt{n_1\mu_1^2+n_2}}\right]_{1\times n_1},\left[\frac{1}{\sqrt{n_1\mu_1^2+n_2}}\right]_{1\times n_2}\right),\label{equ:bar_v_1}\\
\bm{v}_2(\bm{B})&=\left(\left[\frac{\mu_2}{\sqrt{n_1\mu_2^2+n_2}}\right]_{1\times n_1},\left[\frac{1}{\sqrt{n_1\mu_2^2+n_2}}\right]_{1\times n_2}\right),\label{equ:bar_v_2}
\end{align}
where
\begin{align}
\mu_1 &=\frac{(n_1-n_2)\alpha+\sqrt{(n_1-n_2)^2\alpha^2+4n_1n_2\beta^2}}{2n_1\beta}\;,\label{equ:mu_1}\\
\mu_2 &=\frac{(n_1-n_2)\alpha-\sqrt{(n_1-n_2)^2\alpha^2+4n_1n_2\beta^2}}{2n_1\beta}\;.\label{equ:mu_2}
\end{align}
Note that $\mu_1>0$ and $\mu_2<0$ for any $n_1$ and $n_2$, so $\lambda_1(\bm{B})>\lambda_2(\bm{B})$. Moreover, due to $\mu_2<0$, the first $n_1$ entries in $\bm{v}_2(\bm{B})$ have opposite signs compared to the last $n_2$ ones. Based on this observation, if we partition the indices into two sets with respect to their signs in $\bm{v}_2(\bm{B})$, the result naturally provides a classification corresponding to different blocks of matrix $\bm{B}$.

\subsubsection{Adjacent Matrix Generated from Planted Model}
Till now, we have introduced the intuition for performing partitioning on the planted model, i.e., the second eigenvector has the inherent ability to distinguish different block indices. The next step is to relate the planted model $\bm{B}$ to the adjacent matrix $\bm{A}$, as constructed by \eqref{equ:Majority_Voting} in the algorithm. Note that each entry in $\bm{A}$ is obtained from a majority voting among all random sequenced reads covering the corresponding SNP sites. Hence, entries in the upper triangle matrix of $\bm{A}$ are random and independent. Moreover, the distribution of each entry is Bernoulli, and its parameter only depends on whether the corresponding SNP sites are from the same block or not. To this end, two parameters are enough to characterize the distribution of $\bm{A}$, and this provides an opportunity to connect $\bm{A}$ to $\bm{B}$, with respect to a proper permutation of rows and columns (note that permutation does not influence the eigenvectors). In particular, for any $(u,v)\in\{1,\ldots,n\}\times\{1,\ldots,n\}$ with $u> v$, we define
\begin{align}
&\text{Pr}\{a_{uv}=1\}=\pi(b_{uv}),\nonumber\\
&\text{Pr}\{a_{uv}=0\}=1-\pi(b_{uv}),\nonumber
\end{align}
where $\pi$ is the permutation of rows and columns. In words, $\alpha$ is the probability that two SNP sites from the same cluster are detected correctly by majority voting, while $\beta$ is the probability that two SNP sites from different clusters are detected inaccurately. Evidently, $\alpha$ and $\beta$ are closely related to the sequencing techniques, more precisely, the parameters $n$, $m$, and $p$. In our case of unreliable paired-end sequencing, the explicit ways of calculating $\alpha$ and $\beta$ are described as follows:
\begin{align}
\alpha  &\triangleq \text{Pr}\{\text{majority voting claims $a_{uv}=1$}|h_u=h_v\}\nonumber\\
        &=\sum_{i=1}^{m}\text{Pr}\{\text{majority voting claims $a_{uv}=1$, $i$ reads cover SNP sites $u$ and $v$}|h_u=h_v\}\nonumber\\
        &=\sum_{i=1}^m \left\{{m\choose i}\left[\frac{2}{n(n-1)}\right]^i\left[1-\frac{2}{n(n-1)}\right]^{m-i}\sum_{l=\lfloor i/2\rfloor+1}^i{i\choose l}[(1-p)^2+p^2]^l[2p(1-p)]^{i-l}\right\},\nonumber
\end{align}
where $2/n(n-1)$ is the probability of one particular read covering target SNP sites $u$ and $v$; $(1-p)^2+p^2$ is the probability that a particular read observes the SNPs are the identical given the fact $h_u=h_v$; the second summation ranging from $\lfloor i/2\rfloor+1$ to $i$ represents for the majority voting among $i$ voters. Analogously, we have
\begin{align}
\beta  &\triangleq \text{Pr}\{\text{majority voting claims $a_{uv}=1$}|h_u\neq h_v\}\nonumber\\
&=\sum_{i=1}^m \left\{{m\choose i}\left[\frac{2}{n(n-1)}\right]^i\left[1-\frac{2}{n(n-1)}\right]^{m-i}\sum_{l=\lfloor i/2\rfloor+1}^i{i\choose l}[2p(1-p)]^l[(1-p)^2+p^2]^{i-l}\right\},\nonumber
\end{align}
where $2p(1-p)$ is the probability that a particular read observes the SNPs are identical given the fact $h_u\neq h_v$.
Neither $\alpha$ or $\beta$ is straightforward to calculate, however, we could still seek a lower bound for $\alpha$ and an upper bound for $\beta$, to make a worse case discussion.
\begin{lemma}{\label{lem:bounds}}
For the choice of $m=\Theta(n\ln n)$ with sufficient large $n$, there exist positive constants $\kappa_1$, $\kappa_2$, and $\kappa_3$, such that
\begin{align}
\alpha &\geq \frac{2\kappa_1\kappa_2[(1-p)^2+p^2]\ln n}{n-1},\label{equ:alpha_bound}\\
\beta  &\leq \frac{2\kappa_1[2p(1-p)]\ln n}{(n-1)(1-\kappa_3^{-1})},\label{equ:nbeta_bound}
\end{align}
where $\kappa_2<1$ and $\kappa_3>1$.
\end{lemma}
This lemma shows both $\alpha$ and $\beta$ have $\Theta(n^{-1}\ln n)$ scale bounds, and the proof is left in Appendix~\ref{app:lemma1}. Next, using this scale bound, we are ready to show the differences between eigenvectors of $\bm{A}$ and $\bm{B}$ are small, such that their signs are identical with high probability.

\subsubsection{Matrix Eigenvector Perturbation}
After revealing the relationship between adjacent matrix $\bm{A}$ and planted model $\bm{B}$, we move on to discover the difference between their eigenvectors, using matrix perturbation theory. To this end, we show that for our choices of $\alpha$ and $\beta$, the second eigenvector of $\bm{A}$ has a vanishing perturbation compared to the one of $\bm{B}$. This observation provides a theoretical base to perform spectral partitioning on $\bm{A}$, instead of on $\bm{B}$, without much loss in performance.

The classical matrix perturbation theory allows one to determine the sensitivity of the matrix eigenvalues and eigenvectors with respect to a slight influence. \cite{furedi1981eigenvalues} pioneers this
area by providing a general bound for matrix eigenvalue perturbation, and the later work \cite{vu2005spectral} improves this bound, under further assumption to the matrices. Meanwhile, the famous Davis-Kahan sin-theta theorem \cite{davis1970rotation} characterizes the rotation of eigenvectors after perturbation, and \cite{vu2011singular} focuses on random matrices to propose a probabilistic sin-theta theorem. Compared to those general results, the observed matrices in haplotype assembly problem always contain particular structures, for instance, independent and binary distributed entries, low rank, and etc. To this end, we follow the result from a recent perturbation study \cite{tomozei2010distributed}, which has a much tighter bound with respect to binary random matrices, summarized in a lemma as follows.
\begin{lemma}[Lemma 2 and 3 in \cite{tomozei2010distributed}]\label{lem:perturbation}
Consider a square $n\times n$ symmetric $0$-diagonal random matrix $\bm{M}$ such that its elements $m_{uv}=m_{vu}$ are independent Bernoulli random variables with parameters $\mathbb{E}[m_{uv}]=\rho_{uv}\chi n^{-1}$, where $\rho_{uv}$ are constants and $\chi=\Omega(\ln n)$. Then, with probability at least $1-O(n^{-1})$, we have
\begin{align}
&|\lambda_k(\bm{M})-\lambda_k(\mathbb{E}[\bm{M}])|\leq O(\chi^{1/2}),\label{equ:thm_lambda}\\
&||\bm{v}_k(\bm{M})-\bm{v}_k(\mathbb{E}[\bm{M}])||\leq O(\chi^{-1/2}),\label{equ:thm_v}
\end{align}
for any $k$ not larger than the rank of $\mathbb{E}[\bm{M}]$, where $\lambda_k(\bm{M})$ is the $k$-th largest eigenvalue of $\bm{M}$, and $\bm{v}_k(\bm{M})$ is the corresponding $k$-th eigenvector.
\end{lemma}

Back to the haplotype assembly problem, we utilize the lemma directly by considering $\bm{M}$ as the adjacent matrix $\bm{A}$. Note that $\bm{A}$ is already a $0$-diagonal random matrix, with each entry independently distributed as Bernoulli random variable. The parameters of these Bernoulli distributions, i.e., $\alpha$ and $\beta$, satisfy the scale constraints with $\chi=\ln n$, due to Lemma~\ref{lem:bounds}. Moreover, note that $\mathbb{E}[\bm{A}]=\pi(\tilde{\bm{B}})$, where $\tilde{\bm{B}}=\bm{B}-\alpha\bm{I}$, and permutation $\pi$ does not change the eigenvectors. Hence, from \eqref{equ:thm_v}, we have
\begin{align}
||\bm{v}_2(\bm{A})-\bm{v}_2(\tilde{\bm{B}})||\leq O(\ln ^{-1/2}n).\nonumber
\end{align}
By noting that adding an identity matrix does not influence the eigenvectors either, we have $\bm{v}_2(\tilde{\bm{B}})=\bm{v}_2(\bm{B})$. Thus, we obtain
\begin{align}
||\bm{v}_2(\bm{A})-\bm{v}_2(\bm{B})||\leq O(\ln ^{-1/2}n).\label{equ:thm_fraction}
\end{align}
Recall that $\bm{v}_2(\bm{B})$ has a particular form of \eqref{equ:bar_v_2}, which implies that a particular entry perturbed to change its sign should at least contribute $\Omega(n^{-1/2})$ to $||\bm{v}_2(\bm{A})-\bm{v}_2(\bm{B})||$. To this end, if $n_e$ number of errors happen, we have
\begin{align}
\sqrt{\frac{n_e}{n}}\leq O(\ln ^{-1/2}n).
\end{align}
By noting that $n_e/n$ is the fraction of partition errors, we conclude the original haplotype can be recovered reliably with vanishing fraction of errors for large enough $n$.

\begin{remark}
As indicated in the analysis, spectral partitioning using SVD technique could only guarantee the fraction of partition errors vanishes with high probability. For a stronger argument, i.e., the probability of existing partition error tends to zero, one may refer to another technique, ``combinational projection'' \cite{mcsherry2001spectral}, instead of performing SVD directly. Essentially, combinational projection gives another projection, after the one to singular space, onto the span of characteristic vectors generated from particular threshold. In this way, the variances of target random variables are significantly reduced such that Chernoff-type argument could be adopted for a tighter bound on the distance of row spaces after projection. Note that \eqref{equ:thm_lambda} still holds in this case, and by replacing the corresponding bounds in \cite{mcsherry2001spectral}, $\Theta(n \ln n)$ number of reads is sufficient to exactly recover the original haplotype with high probability.
\end{remark}

\begin{remark}
Spectral partitioning is such a simple and efficient algorithm that only majority voting and SVD technique are required for haplotype assembly. In fact, we even do not need a full SVD calculation, since only the second eigenvector is enough to determine the haplotype, as described in the algorithm. To this end, by using power method to discover eigenvector, especially on the parse adjacent matrix (the number of total entries observed is roughly $n\ln n$), the complexity of spectral partitioning can be reduced from the scale of $O(n^3)$ for general cases to $O(n\ln n)$ for our case.
\end{remark}

\begin{remark}
An alternative low-rank matrix interpretation for quantifying the minimum number of entries needed to recover a low rank matrix from observation matrix directly is based on optimality. In this branch of work \cite{candes2010power} \cite{recht2011simpler} \cite{wright2009robust}, an optimization approach is utilized to determine the necessary conditions, and the recovery is facilitated by solving the resulting convex program. In particular, for haplotype assembly, the observed fragments matrix $\bm{R}$ could be considered as a combination of the true haplotype matrix $\bm{S}$ with an independent sequencing error matrix $\bm{N}$. Then, MEC criterion translates into the minimum $l_1$-norm of $\bm{N}$, and the optimization problem is given by
\begin{align}
\textrm{min}  &\quad||\bm{S}||_*+\gamma ||\bm{N}||_1\nonumber\\
\textrm{s.t.} &\quad\mathcal{P}_{\Omega}(\bm{S}+\bm{N})=\mathcal{P}_{\Omega}(\bm{R}),\nonumber
\end{align}
where $\|\bm{S}\|_*$ is the nuclear norm of $\bm{S}$, and $\gamma$ is the balancing weight. \cite{candes2011robust} \cite{chen2011low} report that the row space of original matrix could be reliably recovered, as long as the number of observed entries is large enough. Putting it more precisely, the number of reads needed for recovery is as least $\Omega(n\cdot\textrm{poly}(\ln n))$, which does not outperform the bound from spectral partitioning. The kernel technique utilized in general for this type of proofs is Golfing Scheme \cite{candes2011robust} \cite{chen2011low}, which requires a lower bound on the number of sampled entries to construct dual certificate. If new scheme could be found to replace Golfing Scheme with better performance guarantee, especially for the particular structure of haplotyping problem, then, optimality method may also meet the necessary condition with only a log-factor gap.
\end{remark}


\section{Simulation Results and Analyses}\label{sec:Simulation}

\subsection{Testification on Simulated Database}
In this part, we testify the performance of the two proposed algorithms, erasure decoding and spectral partitioning, on an ideal simulated database. In particular, in this database, haplotypes are randomly generated according to uniform distribution, then paired-end fragments are sampled from haplotypes randomly and uniformly, with i.i.d. sampling errors. For the moment, we fix the number of SNPs observed in each fragment as $2$. The target of this simulation is to study the relations among three key parameters in the algorithms, i.e., the length of haplotype $n$, the probability of sampling errors $p$, and most importantly, the number of sampled reads $m$. We show that the simulation results verify the conclusions of theorems in this paper, and also provide an intuition for choosing proper parameters from the practical perspective.

\begin{figure}[h]
  \centering
  \includegraphics[width=1\columnwidth]{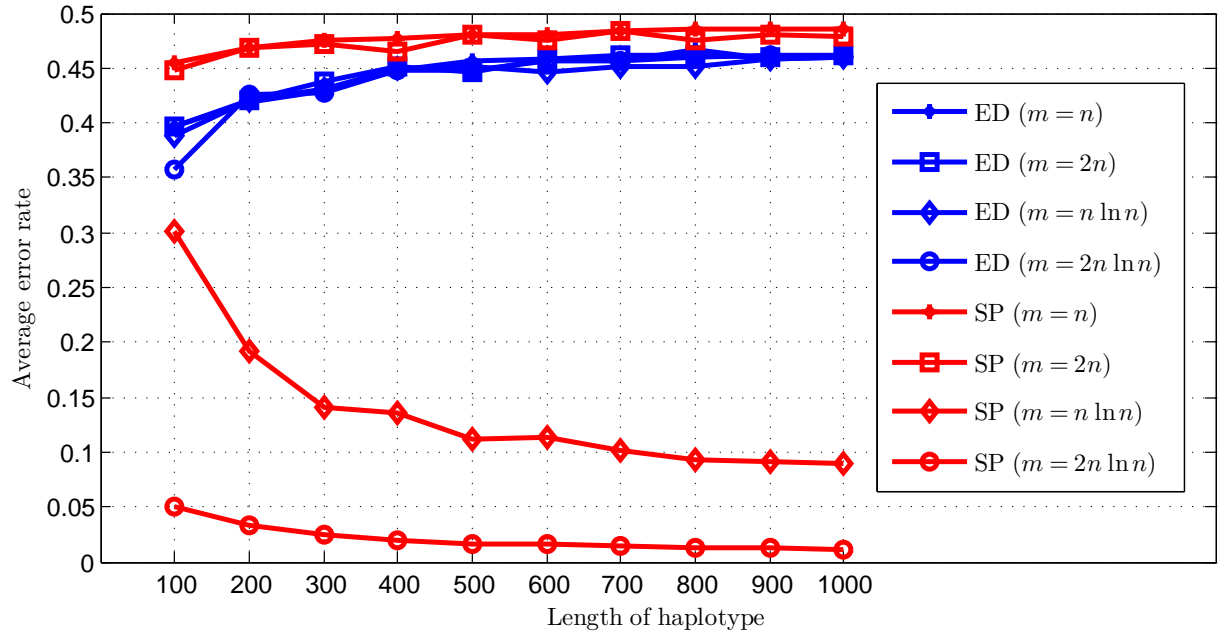}
  \caption{Plot of average error rates from $100$ random simulations by fixing the probability of sampling of errors as $p=0.1$. In this simulation, we focus on different scale choices for the number of reads $m$, with respect to both of the algorithms, erasure decoding (ED) and spectral partitioning (SP).}\label{fig:Changing_Reads}
\end{figure}

To start with, we fix the probability of sampling error as $p=0.1$ (much larger than the typical value in practice), and study the relation between the scale choices of $m$ and different lengths of haplotypes $n$. We plot the simulation results in Fig.~\ref{fig:Changing_Reads}, and provide the following observations.
\begin{itemize}
\item Erasure decoding algorithm fails to recover for all choices of $m$, which is basically due to large sampling noises. As indicated in the theoretical analysis, this algorithm is intuitively designed for the error-free case, and it has no performance guarantee when adopted for erroneous case.
\item For spectral partitioning, choosing $m=\Theta(n)$ is not enough for reliable recovery, while choosing $m=\Theta(n\ln n)$ is sufficient to make error fraction vanish with large enough $n$. This result is consistent with the conclusion in Theorem~\ref{thm:SP}.
\item Spectral partitioning, when implemented with large enough number of reads (i.e., $m=\Theta(n \ln n)$), has better error rate for larger length of haplotype. \eqref{equ:thm_fraction} provides a theoretical foundation at this point.
\end{itemize}

\begin{figure}[h]
  \centering
  \includegraphics[width=1\columnwidth]{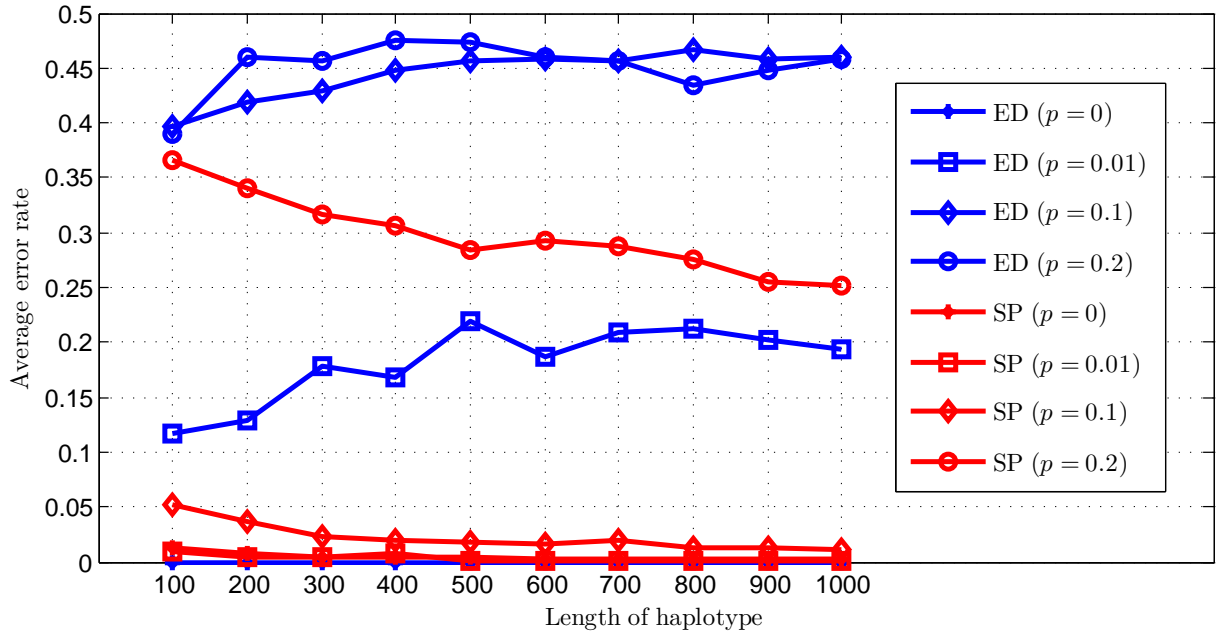}
  \caption{Plot of average error rates from $100$ random simulations by fixing the number of reads as $m=2n\ln n$. In this simulation, we focus on the performance under different choice of sampling errors, with respect to both of the algorithms, erasure decoding (ED) and spectral partitioning (SP).}\label{fig:Changing_Error}
\end{figure}

Then, in the light of the theoretical analysis and the previous simulation results, we fix the number of sampled reads as $m=2n\ln n$, and study the performance for both algorithms under different probabilities of sampling errors, with respect to diverse lengths of haplotypes. Simulation results are illustrated in Fig.~\ref{fig:Changing_Error}, and show the following observations.
\begin{itemize}
\item Erasure decoding algorithm works extremely well for error-free case, when implemented with large enough number of fragments. However, for erroneous case, this algorithm fails to recover the original haplotypes with high confidence.
\item The convergence rate for spectral partitioning highly depends on $p$. In other words, spectral partitioning is more proper to be utilized in low-noise case, i.e., $p\leq0.1$, which already covers most of the practical applications.
\end{itemize}

These two simple simulations on ideal simulated database testify our theoretical analyses, and we summarize the performances of these two algorithms proposed in this paper.
\begin{enumerate}
\item Erasure decoding only fits for the noise-free case, and it requires the size of fragment assembly to be at least scaling as $m=\Theta(n\ln n)$ for reliable recovery.
\item Spectral partitioning fits for the low-noise case, i.e., $p\leq 0.1$, and it also requires the number of reads to scale as $m=\Theta(n\ln n)$. When these two conditions are satisfied, spectral partitioning is capable of recovering the original haplotype with high confidence, and the recovery rate decreases when the length of haplotype grows.
\end{enumerate}

\subsection{Simulation on Benchmark Database}

In this part, we study the performance of both algorithms on the database created by \cite{geraci2010comparison}, which is generated from the Phase I of HapMap project \cite{international2005haplotype}, and widely adopted as a benchmark for evaluating the effectiveness of algorithms. This database consists of all $22$ chromosomes from $209$ unrelated individuals, and shotgun sequencing process has been simulated to obtain the SNP observation matrix. Note that only heterogeneous SNP sites are considered in our simulation, and the recovery rate is based on the length after filtering. Moreover, in every fragment, the number of SNPs covered is not fixed to $2$ in this database, however, our algorithms can still be utilized directly on this database, since the $2$ observations assumption is only for theoretical analysis and the algorithms themselves are not restricted to that case.

TABLE~\ref{tab:simulation} shows the average recovery rate over $100$ randomly generated data set from \cite{geraci2010comparison}, where free parameters includes: 1) the haplotype length $n=100, 350, 700$; 2) the coverage $c=3,5,8,10$; and 3) the sampling error rate $p=0\%, 10\%, 20\%$. From these simulation results, we find erasure decoding successfully recover original haplotype with high probability when $p=0$, but fails when $p\neq 0$. Moreover, sparse partitioning performs well, at least comparable with the existing algorithms, when implemented with large number of sampled reads. To this end, our proposed algorithms, which are initially designed from the theoretically optimal perspective, also have practical significance.
\begin{table}[!h]
\begin{tabular}{c|l|cccc|cccc|cccc}
\toprule[2pt]
\multicolumn{2}{c|}{\multirow{2}*{\textbf{Algorithms}}}  & \multicolumn{4}{c}{$p=0.0$} & \multicolumn{4}{|c}{$p=0.1$} & \multicolumn{4}{|c}{$p=0.2$} \\
                                \cmidrule{3-14}
\multicolumn{2}{c|}{}& $c=3$ & $c=5$ & $c=8$ & $c=10$  & $c=3$ & $c=5$ & $c=8$ & $c=10$ & $c=3$ & $c=5$ & $c=8$ & $c=10$\\
\midrule[1pt]

\multirow{9}{*}{\begin{turn}{90}$n=100$\end{turn}}
  &SpeedHap  & $0.999$ & $1.000$ & $1.000$ & $1.000$ & $0.895$ & $0.967$ & $0.989$ & $0.990$ & $0.623$ & $0.799$ & $0.852$ & $0.865$ \\
  &Fast Hare & $0.999$ & $0.999$ & $1.000$ & $1.000$ & $0.919$ & $0.965$ & $0.993$ & $0.998$ & $0.715$ & $0.797$ & $0.881$ & $0.915$ \\
  &2d-mec    & $0.990$ & $0.997$ & $1.000$ & $1.000$ & $0.912$ & $0.951$ & $0.983$ & $0.988$ & $0.738$ & $0.793$ & $0.873$ & $0.894$ \\
  &HapCUT    & $1.000$ & $1.000$ & $1.000$ & $1.000$ & $0.929$ & $0.920$ & $0.901$ & $0.892$ & $0.782$ & $0.838$ & $0.864$ & $0.871$ \\
  &MLF       & $0.973$ & $0.992$ & $0.997$ & $0.998$ & $0.889$ & $0.970$ & $0.985$ & $0.995$ & $0.725$ & $0.836$ & $0.918$ & $0.938$ \\
  &SHR-three & $0.816$ & $0.861$ & $0.912$ & $0.944$ & $0.696$ & $0.738$ & $0.758$ & $0.762$ & $0.615$ & $0.655$ & $0.681$ & $0.699$ \\
  &DGS       & $1.000$ & $1.000$ & $1.000$ & $1.000$ & $0.930$ & $0.985$ & $0.989$ & $0.997$ & $0.725$ & $0.813$ & $0.878$ & $0.917$ \\
  &{\color{blue}ED}        & {\color{blue}$1.000$} & {\color{blue}$1.000$} & {\color{blue}$1.000$} & {\color{blue}$1.000$} & {\color{blue}$0.650$} & {\color{blue}$0.651$} & {\color{blue}$0.627$} & {\color{blue}$0.639$} & {\color{blue}$0.587$} & {\color{blue}$0.581$} & {\color{blue}$0.585$} & {\color{blue}$0.593$} \\
  &{\color{blue}SP}        & {\color{blue}$0.958$} & {\color{blue}$0.997$} & {\color{blue}$0.999$} & {\color{blue}$1.000$} & {\color{blue}$0.883$} & {\color{blue}$0.961$} & {\color{blue}$0.990$} & {\color{blue}$0.995$} & {\color{blue}$0.687$} & {\color{blue}$0.809$} & {\color{blue}$0.918$} & {\color{blue}$0.943$} \\
\midrule[1pt]
\multirow{9}{*}{\begin{turn}{90}$n=350$\end{turn}}
  &SpeedHap  & $0.999$ & $1.000$ & $1.000$ & $1.000$ & $0.819$ & $0.959$ & $0.984$ & $0.984$ & $0.439$ & $0.729$ & $0.825$ & $0.855$ \\
  &Fast Hare & $0.990$ & $0.999$ & $1.000$ & $0.999$ & $0.871$ & $0.945$ & $0.985$ & $0.995$ & $0.684$ & $0.746$ & $0.853$ & $0.877$ \\
  &2d-mec    & $0.965$ & $0.993$ & $0.998$ & $0.999$ & $0.837$ & $0.913$ & $0.964$ & $0.978$ & $0.675$ & $0.729$ & $0.791$ & $0.817$ \\
  &HapCUT    & $1.000$ & $1.000$ & $1.000$ & $1.000$ & $0.930$ & $0.913$ & $0.896$ & $0.888$ & $0.771$ & $0.831$ & $0.862$ & $0.867$ \\
  &MLF       & $0.864$ & $0.929$ & $0.969$ & $0.981$ & $0.752$ & $0.858$ & $0.933$ & $0.962$ & $0.642$ & $0.728$ & $0.798$ & $0.831$ \\
  &SHR-three & $0.830$ & $0.829$ & $0.895$ & $0.878$ & $0.682$ & $0.724$ & $0.742$ & $0.728$ & $0.591$ & $0.632$ & $0.670$ & $0.668$ \\
  &DGS       & $0.999$ & $1.000$ & $1.000$ & $1.000$ & $0.926$ & $0.978$ & $0.996$ & $0.998$ & $0.691$ & $0.769$ & $0.842$ & $0.878$ \\
  &{\color{blue}ED}        & {\color{blue}$1.000$} & {\color{blue}$1.000$} & {\color{blue}$1.000$} & {\color{blue}$1.000$} & {\color{blue}$0.608$} & {\color{blue}$0.595$} & {\color{blue}$0.587$} & {\color{blue}$0.586$} & {\color{blue}$0.553$} & {\color{blue}$0.549$} & {\color{blue}$0.538$} & {\color{blue}$0.547$} \\
  &{\color{blue}SP}        & {\color{blue}$0.903$} & {\color{blue}$0.972$} & {\color{blue}$0.992$} & {\color{blue}$0.997$} & {\color{blue}$0.768$} & {\color{blue}$0.933$} & {\color{blue}$0.983$} & {\color{blue}$0.992$} & {\color{blue}$0.598$} & {\color{blue}$0.679$} & {\color{blue}$0.843$} & {\color{blue}$0.905$} \\
\midrule[1pt]
\multirow{9}{*}{\begin{turn}{90}$n=700$\end{turn}}
  &SpeedHap  & $0.999$ & $1.000$ & $1.000$ & $1.000$ & $0.705$ & $0.947$ & $0.985$ & $0.986$ & $0.199$ & $0.681$ & $0.801$ & $0.813$ \\
  &Fast Hare & $0.988$ & $0.999$ & $1.000$ & $0.999$ & $0.829$ & $0.949$ & $0.986$ & $0.995$ & $0.652$ & $0.712$ & $0.808$ & $0.872$ \\
  &2d-mec    & $0.946$ & $0.976$ & $0.992$ & $0.997$ & $0.786$ & $0.880$ & $0.948$ & $0.965$ & $0.647$ & $0.697$ & $0.751$ & $0.778$ \\
  &HapCUT    & $1.000$ & $1.000$ & $1.000$ & $1.000$ & $0.927$ & $0.916$ & $0.896$ & $0.889$ & $0.753$ & $0.825$ & $0.856$ & $0.861$ \\
  &MLF       & $0.787$ & $0.854$ & $0.919$ & $0.933$ & $0.698$ & $0.809$ & $0.863$ & $0.884$ & $0.624$ & $0.682$ & $0.747$ & $0.765$ \\
  &SHR-three & $0.781$ & $0.832$ & $0.868$ & $0.898$ & $0.668$ & $0.716$ & $0.743$ & $0.726$ & $0.591$ & $0.617$ & $0.653$ & $0.675$ \\
  &DGS       & $0.999$ & $1.000$ & $1.000$ & $1.000$ & $0.931$ & $0.977$ & $0.987$ & $0.997$ & $0.669$ & $0.741$ & $0.818$ & $0.861$ \\
  &{\color{blue}ED}        & {\color{blue}$1.000$} & {\color{blue}$1.000$} & {\color{blue}$1.000$} & {\color{blue}$1.000$} & {\color{blue}$0.576$} & {\color{blue}$0.571$} & {\color{blue}$0.572$} & {\color{blue}$0.573$} & {\color{blue}$0.534$} & {\color{blue}$0.532$} & {\color{blue}$0.531$} & {\color{blue}$0.528$} \\
  &{\color{blue}SP}        & {\color{blue}$0.887$} & {\color{blue}$0.967$} & {\color{blue}$0.991$} & {\color{blue}$0.997$} & {\color{blue}$0.723$} & {\color{blue}$0.910$} & {\color{blue}$0.977$} & {\color{blue}$0.990$} & {\color{blue}$0.562$} & {\color{blue}$0.610$} & {\color{blue}$0.751$} & {\color{blue}$0.843$} \\
\bottomrule[2pt]
\end{tabular}
\caption{Comparisons of our algorithms, erasure decoding (ED) and spectral partitioning (SP), with existing algorithms listed in [22]. Each entry in the table represents the average recovery rate from $100$ randomly generated haplotype observation matrices, with respect to different $n$, $c$, and $p$.}
\label{tab:simulation}
\end{table}


\section{Conclusion}\label{sec:Conclusion}

In this paper, we consider the haplotype assembly problem from an information theoretic perspective.
To determine chromosome membership of reads provided by high-throughput sequencing systems
and to reconstruct haplotypes, we consider them as messages that are encoded and transmitted over
particular channel model. This channel model reflects the salient features of the paired-end sequencing
technology, e.g., that every row in a data matrix contains few randomly placed entries.

In the case of error-free sequencing, we show that the required number of reads needed for reconstruction is at least
of the same order as the length of the haplotype sequence. The proof basically traces classical information theory analysis on channel capacity.
From the perspective of sufficient condition, the erasure decoding algorithm utilizes the common information across rows to iteratively recover haplotypes.
We show that this algorithm ensures reconstruction with the optimal scale of reads, regardless of a log-factor gap.

In the case of erroneous sequencing, where errors are assumed to be generated independently and identically, we show that the necessary condition
for the number of reads to recover the haplotypes is analogue to error-free case, i.e., the same scale of haplotype length. To study the sufficient condition,
we reshape the original haplotyping problem into a low-rank matrix interpretation. Using matrix permutation theory, we illustrate that haplotype sequences could be recovered reliably, when implemented with $\Theta(n\ln n)$ number of fragments.

Both theoretical analyses and simulation results provide support to the conclusion in this paper, and the information theoretic view of regarding a haplotype assembly problem could be generalized to more applications.


\section*{Acknowledgements}

The authors would like to thank Yudong Chen, Xinyang Yi, and Soyeon Ahn for insightful discussions and comments.

\bibliographystyle{IEEEtran}

\appendices

\section{Proof to Lemma ~\ref{lem:bounds}}\label{app:lemma1}
Assume $m=\kappa_1 n\ln n$, where $\kappa_1$ is a positive constant. In order to provide a lower bound for $\alpha$, we truncate the first summation by leaving only the term with $i=1$. More precisely, we have
\begin{align}
\alpha  &=\sum_{i=1}^m \left\{{m\choose i}\left[\frac{2}{n(n-1)}\right]^i\left[1-\frac{2}{n(n-1)}\right]^{m-i}\sum_{l=\lfloor i/2\rfloor+1}^i{i\choose l}[(1-p)^2+p^2]^l[2p(1-p)]^{i-l}\right\}\nonumber\\
        &\geq {m\choose 1}\left[\frac{2}{n(n-1)}\right]\left[1-\frac{2}{n(n-1)}\right]^{m-1}{1\choose 1}[(1-p)^2+p^2][2p(1-p)]^0\nonumber\\
        &\geq \frac{2\kappa_1n\ln n}{n(n-1)} e^{-\frac{4\kappa_1n\ln n}{n(n-1)}} [(1-p)^2+p^2]\nonumber\\
        &=\frac{2\kappa_1[(1-p)^2+p^2]n^{-\frac{4\kappa_1}{n-1}}\ln n}{n-1}.\nonumber
\end{align}
Note that $n^{-\frac{4\kappa_1}{n-1}}$ is an increasing function with $n$ and tends to $1$. Hence, for large enough $n$, there exists a constant $\kappa_2<1$ such that
\begin{align}
n^{-\frac{4\kappa_1}{n-1}}\geq\kappa_2. \label{fun:lem_1_assume1}
\end{align}
To this end, the lower bound becomes
\begin{align}
\alpha \geq \frac{2\kappa_1\kappa_2[(1-p)^2+p^2]\ln n}{n-1}.\label{fun:lem_1_alpha}
\end{align}
Thus, $\alpha$ has a $\Theta(n^{-1}\ln n)$ scale lower bound. In fact, this bound is quite tight, because the first term ($i=1$) dominates the value (analogue to the analysis of $\beta$).

On the other hand, we aim to propose an upper bound for $\beta$. In particular, we show that the terms in summation are at least exponentially decreasing, such that the first term dominates the value of $\beta$. For this purpose, we denote
\begin{align}
\beta_i\triangleq {m\choose i}\left[\frac{2}{n(n-1)}\right]^i\left[1-\frac{2}{n(n-1)}\right]^{m-i}\sum_{l=\lfloor i/2\rfloor+1}^i{i\choose l}[2p(1-p)]^l[(1-p)^2+p^2]^{i-l},\nonumber
\end{align}
and
\begin{align}
\beta_i^{(l)}\triangleq {i\choose l}[2p(1-p)]^l[(1-p)^2+p^2]^{i-l},\nonumber
\end{align}
then
$$\beta=\sum\limits_{i=1}^m\beta_i,$$
and
$$\beta_i={m\choose i}\left[\frac{2}{n(n-1)}\right]^i\left[1-\frac{2}{n(n-1)}\right]^{m-i}\sum_{l=\lfloor i/2\rfloor+1}^i\beta_i^{(l)}.$$
In order to derive a lower bound on $\beta_i/\beta_{i+1}$ for any $i$, we focus on two cases:
\begin{enumerate}
\item For $i$ even, assume $i=2k$, then
\begin{align}
\frac{\beta_{2k}}{\beta_{2k+1}}
&=\frac{{m\choose 2k}\left[\frac{2}{n(n-1)}\right]^{2k}\left[1-\frac{2}{n(n-1)}\right]^{m-2k}\sum\limits_{l=k+1}^{2k}\beta_{2k}^{(l)}}
{{m\choose 2k+1}\left[\frac{2}{n(n-1)}\right]^{2k+1}\left[1-\frac{2}{n(n-1)}\right]^{m-2k-1}\sum\limits_{l=k+1}^{2k+1}\beta_{2k+1}^{(l)}}\nonumber\\
&=\frac{(2k+1)[n(n-1)-2]\sum\limits_{l=k+1}^{2k}\beta_{2k}^{(l)}}{2(\kappa_1n\ln n-2k)\sum\limits_{l=k+1}^{2k+1}\beta_{2k+1}^{(l)}}.\nonumber
\end{align}
Note that there are $k+1$ terms for $\beta_{2k+1}^{(l)}$ in the denominator, but only $k$ terms for $\beta_{2k}^{(l)}$ in the numerator.
Hence, we duplicate the numerator to compare with denominator. More precisely, for $k+1\leq l\leq 2k$,
\begin{align}
\frac{\beta_{2k}^{(l)}}{\beta_{2k+1}^{(l)}}=\frac{2k+1-l}{(2k+1)[(1-p)^2+p^2]}\geq \frac{1}{2k+1},\label{fun:lem_1_piece1}
\end{align}
where the last inequality holds due to $(1-p)^2+p^2\leq 1$. Moreover,
\begin{align}
\frac{\beta_{2k}^{(k+1)}}{\beta_{2k+1}^{(2k+1)}}=\frac{(2k)![(1-p)^2+p^2]^{k-1}}{(k+1)!(k-1)![2p(1-p)]^k}\geq \frac{1}{2k+1},\label{fun:lem_1_piece2}
\end{align}
where the last inequality holds due to $1\geq(1-p)^2+p^2\geq 2p(1-p)$. Combining these two pieces together, we have
\begin{align}
\frac{2\beta_{2k}}{\beta_{2k+1}}
&=\frac{(2k+1)[n(n-1)-2]\left\{\sum\limits_{l=k+1}^{2k}\beta_{2k}^{(l)}+\sum\limits_{l=k+1}^{2k}\beta_{2k}^{(l)}\right\}}
{2(\kappa_1n\ln n-2k)\left\{\sum\limits_{l=k+1}^{2k}\beta_{2k+1}^{(l)}+\beta_{2k+1}^{(2k+1)}\right\}}\nonumber\\
&\geq\frac{(2k+1)[n(n-1)-2]\left\{\sum\limits_{l=k+1}^{2k}\beta_{2k}^{(l)}+\beta_{2k}^{(k+1)}\right\}}
{2(\kappa_1n\ln n-2k)\left\{\sum\limits_{l=k+1}^{2k}\beta_{2k+1}^{(l)}+\beta_{2k+1}^{(2k+1)}\right\}}\nonumber\\
&\geq \frac{(2k+1)[n(n-1)-2]}{2(\kappa_1n\ln n-2k)(2k+1)}\nonumber\\
&=\frac{n(n-1)-2}{2(\kappa_1n\ln n-2)}.\nonumber
\end{align}
Thus,
\begin{align}
\frac{\beta_{2k}}{\beta_{2k+1}}\geq\frac{n(n-1)-2}{4(\kappa_1n\ln n-2)}.\label{fun:lem_1_bound_even}
\end{align}
\item For $i$ odd, assume $i=2k-1$, then
\begin{align}
\frac{\beta_{2k-1}}{\beta_{2k}}
&=\frac{{m\choose 2k-1}\left[\frac{2}{n(n-1)}\right]^{2k-1}\left[1-\frac{2}{n(n-1)}\right]^{m-2k+1}\sum\limits_{l=k}^{2k-1}\beta_{2k-1}^{(l)}}
{{m\choose 2k}\left[\frac{2}{n(n-1)}\right]^{2k}\left[1-\frac{2}{n(n-1)}\right]^{m-2k}\sum\limits_{l=k+1}^{2k}\beta_{2k}^{(l)}}\nonumber\\
&=\frac{2k[n(n-1)-2]\sum\limits_{l=k}^{2k-1}\beta_{2k-1}^{(l)}}{2(\kappa_1n\ln n-2k+1)\sum\limits_{l=k+1}^{2k}\beta_{2k}^{(l)}}.\nonumber
\end{align}
In this case, both numerator and denominator have $k$ terms in summation. Hence, by comparing one by one, we have
\begin{align}
\frac{\beta_{2k-1}^{(l)}}{\beta_{2k}^{(l)}}=\frac{2k-l}{2k[(1-p)^2+p^2]}\geq \frac{1}{2k}.\label{fun:lem_1_piece3}
\end{align}
Thus,
\begin{align}
\frac{\beta_{2k-1}}{\beta_{2k}}\geq\frac{n(n-1)-2}{2(\kappa_1n\ln n-1)}.\label{fun:lem_1_bound_odd}
\end{align}
\end{enumerate}
Note that in both cases, the lower bounds \eqref{fun:lem_1_bound_even} and \eqref{fun:lem_1_bound_odd} tends to infinity as $n$ increases. To this end, there exists a constant $\kappa_3>1$, such that for large enough $n$,
\begin{align}
\min\left\{\frac{n(n-1)-2}{4(\kappa_1n\ln n-2)}\;,\;\frac{n(n-1)-2}{2(\kappa_1n\ln n-1)}\right\}\geq\kappa_3,\label{fun:lem_1_assume2}
\end{align}
which further implies for any value of $i$,
\begin{align}
\frac{\beta_{i}}{\beta_{i+1}}\geq\kappa_3.\nonumber
\end{align}
Based on this, we obtain $\beta_{i}\leq \beta_{1}\kappa_3^{1-i}$, and
\begin{align}
\beta_1 &={m\choose 1}\left[\frac{2}{n(n-1)}\right]\left[1-\frac{2}{n(n-1)}\right]^{m-1}{1\choose 1}[2p(1-p)][(1-p)^2+p^2]^0\nonumber\\
        &\leq \frac{2\kappa_1n\ln n}{n(n-1)} e^{-\frac{2\kappa_1n\ln n}{n(n-1)}} [2p(1-p)]\nonumber\\
        &=\frac{2\kappa_1[2p(1-p)]n^{-\frac{2\kappa_1}{n-1}}\ln n}{n-1}\nonumber\\
        &\leq \frac{2\kappa_1[2p(1-p)]\ln n}{n-1},\nonumber
\end{align}
where we have used the fact that $n^{-\frac{2\kappa_1}{n-1}}\leq 1$. Hence, we obtain
\begin{align}
\beta   &=\sum_{i=1}^m\beta_i\nonumber\\
        &\leq \sum_{i=1}^m \beta_1\kappa_3^{1-i}\nonumber\\
        &\leq \frac{2\kappa_1[2p(1-p)]\ln n}{(n-1)(1-\kappa_3^{-1})}.\label{fun:lem_1_beta}
\end{align}
Thus, the upper bound for $\beta$ is also $\Theta(n^{-1}\ln n)$ scale.

A point to clarify is, in this proof, we have argued several times about ``large enough $n$''. One may concern that whether the particular choice of $n$ satisfying proof assumptions could match the practical requirement of haplotype assembly. In fact, for instance, we have $\kappa_1=2$ in the simulation setup, then if simply choosing $\kappa_2=1/2$, and $\kappa_3=2$, the minimum value for $n$ to satisfy both assumptions \eqref{fun:lem_1_assume1} and \eqref{fun:lem_1_assume2} is given by
\begin{align}
n\geq \max\{45,69,28\}=69,\nonumber
\end{align}
which is quite smaller than the commonly accepted value in haplotype assembly. To this end, our bounds hold well from the practical perspective.
\end{document}